\documentstyle[11pt]{article}

\renewcommand{\baselinestretch}{1.5}
\def\be{\begin{eqnarray}}
\def\ee{\end{eqnarray}}

\def\Tr{{\rm Tr}\;}

\def\thefootnote{\fnsymbol{footnote}}
\newcommand{\beq}{\begin{eqnarray}}
\newcommand{\eeq}{\end{eqnarray}}
\newcommand{\beqno}{\begin{eqnarray*}}
\newcommand{\eeqno}{\end{eqnarray*}}
\long\def\beginomit#1\endomit{}
\def\ben{\begin{enumerate}}
\def\een{\end{enumerate}}
\def\bi{\begin{itemize}}
\def\ei{\end{itemize}}

\def\eg{{\it e.g.}}

\def\O{{\cal O}}

\def\Tr{{\mbox{Tr}}}

\renewcommand{\thefootnote}{\#\arabic{footnote}}
\begin{document}

\begin{titlepage}\begin{center}
\hfill{ SNUTP-94-28}

\hfill{hep-ph/9403339}

\hfill{\it April 1, 1994}
\vskip 0.7in
{\LARGE\bf Kaon Condensation in ``Nuclear Star" Matter}
\vskip 0.4in
{\large Chang-Hwan Lee$^a$,  G.E. Brown$^b$ and Mannque Rho$^c$}\\
\vskip 0.1in
{\large a) \it Department of Physics and Center for Theoretical Physics,}\\
{\large \it Seoul National University, Seoul 151-742, Korea}\\
{\large b) \it Department of Physics, State University of New York,} \\
{\large \it Stony Brook, N.Y. 11794, USA. }\\
{\large c) \it Service de Physique Th\'{e}orique, CEA  Saclay}\\
{\large\it 91191 Gif-sur-Yvette Cedex, France}
\vskip 1.2in
{\bf ABSTRACT}\\ \vskip 0.1in
\begin{quotation}
The critical density for kaon condensation in ``nuclear star" matter
is computed up to
two-loop order {\it in medium}
(corresponding to next-to-next-to-leading order in chiral perturbation
theory in free space) with a
heavy-baryon effective chiral Lagrangian whose parameters are determined from
$KN$ scattering and kaonic atom data. To the order considered, the kaon
self-energy has highly non-linear density dependence in dense matter. We find
that the four-Fermi interaction terms in the chiral Lagrangian play an
important role in triggering condensation, predicting for ``natural"
values of the four-Fermi interactions a rather
low critical density, $\rho_c < 4 \rho_0$.
\end{quotation}
\end{center}\end{titlepage}

In a recent paper, Brown and Bethe \cite{brownbethe} suggested that if kaon
condensates develop at relatively low matter density in the collapse of large
stars, then low-mass black holes are more likely to form than neutron stars
of the mass greater than 1.5 times the solar mass $M_\odot$.
Ever since the seminal paper of Kaplan and Nelson\cite{KN},
there have been numerous investigations on
kaon condensation in dense neutron star matter as well as in nuclear matter
based both on effective chiral
Lagrangians\cite{BKR,PW,BKRT,TPL,BLRT} and on phenomenological
off-shell meson-nucleon interactions\cite{YNMK,lutz}. The results have been
quite confusing: While the chiral Lagrangian calculations generally
predict a relatively
low critical density, $\rho_c\sim (2-4) \rho_0$, the phenomenological
approaches
have indicated that a kaon condensation at such a low density
may be incompatible with
kaon-nucleon data and in some versions seem to exclude any condensation at al.
It is now understood \cite{manohar,YKM} that the main difference in the two
approaches lies in terms higher than linear in
density in the energy density of the matter.

In this paper, we report the first higher-order
chiral perturbation calculation of the
critical density with a chiral Lagrangian that when calculated
to one loop order (that is to $\O (Q^2)$ relative to the leading order),
correctly describes s-wave kaon-nucleon amplitude near threshold
 {\it and} that includes four-Fermi interactions
constrained by kaonic atom data. Our prediction for critical density is
$\rho_c\approx (3-4) \rho_0$.

To implement spontaneously broken chiral symmetry in the computation,
we take the Jenkins-Manohar heavy-baryon chiral Lagrangian
\cite{JM} as extended in \cite{LJMR} to $\O (Q^3)$ to describe
s-wave kaon nucleon scattering to one loop order in chiral perturbation theory
(ChPT). In addition to the usual octet and decuplet baryons and the octet
pseudo-Goldstone fields, the $\Lambda (1405)$ was found to figure importantly
in the kaon-nucleon process. This field which provides repulsion at threshold
in
$K^-p$ scattering was introduced in \cite{LJMR} as an elementary field.
By fitting the coefficients of the resulting chiral Lagrangian
by empirical kaon nucleon s-wave scattering data at low energy,
it was shown there that higher order chiral corrections
can systematically be calculated while preserving the ``naturalness" condition
for on-shell scattering amplitudes. By a straightforward off-shell
extension, we have predicted an off-shell kaon-nucleon amplitude that could
be applied to kaonic atom \cite{kaonicatom}
as well as kaon condensation phenomena.
The predicted off-shell amplitude was found to be in fair agreement
with the phenomenological fit obtained by Steiner \cite{Steiner}.
A simple way of understanding the result so obtained is to use the
chiral counting appropriate for the meson-baryon system.
In heavy-baryon formalism (HBF), we can order the relevant
observables as a power series in $Q$, say, $Q^\nu$ where $Q$ is the
characteristic energy or momentum scale we are interested in  and $\nu$ an
integer. Thus to leading order, the
kaon-nucleon amplitude $T^{KN}$ goes as ${\cal O}(Q^1)$, to next order
but involving only tree graphs as ${\cal O}(Q^2)$
and to next-to-next order (or N$^2$LO)
at which one-loop graphs enter as ${\cal O}(Q^3)$. The off-shell amplitude
calculated in \cite{LJMR} contains therefore all terms up to $\O (Q^3)$.
This amplitude could be used in impulse approximation
for in-medium processes. The corresponding contribution to the kaon self-energy
given by Fig.1a is
    \be
    \Pi^{imp}_K(\omega) &=& -\left( \rho_p  {\cal T}^{K^-p}_{free}(\omega)
        +\rho_n  {\cal T}^{K^-n}_{free}(\omega) \right)\label{self1}
    \ee
where ${\cal T}^{KN}$ is the off-shell s-wave KN transition matrix
\footnote{The amplitude ${\cal T}^{KN}$ taken on-shell, {\it i.e.},
$\omega=M_K$, and the scattering length
$a^{KN}$ are related by $ a^{KN} = \frac{1}{4\pi (1+M_K/m_B)}
T^{KN} $.}. This can provide an optical potential for kaonic atom and the
linear density approximation for kaon condensation. We will shortly discuss
what critical density is obtained in this approximation.

To go beyond the linear density approximation
as required for a more reliable treatment of
both kaon condensation and kaonic atom,
we need to compute the {\it effective action} (or {\it effective potential}
for uniform
matter). For this the first obvious correction to the self-energy
(\ref{self1}) comes from the influence of the medium on the amplitude
${\cal T^{KN}}$ which is readily taken into account by replacing
the heavy nucleon propagator in Figs.1b-1f by the in-medium one
    \be
    G^0(k) \simeq \frac{i}{v\cdot k +i\epsilon}
    -2\pi\delta(k_0)\theta (k_F -|\vec k|)
    \label{prop}\ee
where $k_F$ is the nucleon Fermi momentum related to density $\rho_N$ by
the usual relation
    $\rho_N
    =\frac{\gamma}{6\pi^2} k_{F_N}^3$
with the degeneracy factor $\gamma=2$ for neutron and proton in nuclear matter.
We shall call this class of corrections
$\delta {\cal T}_{\rho N}^{K^- N}$. It is clear
that it will give rise to non-linear density dependence. Furthermore
we expect it to be repulsive as it corresponds
to the Pauli exclusion effect.

The second correction -- which is a lot more
important -- comes from ``particle-hole" excitations
that do not figure in $KN$ scattering but can contribute importantly in medium.
These are depicted in Figs.2.
Since we are dealing with s-wave kaon interaction, the most important
configuration that $K^-$ can couple to is the $\Lambda (1405)$ particle-nucleon
hole (denoted as $\Lambda N^{-1}$ with $N$ either a proton ($p$) or neutron
($n$)). Thus Fig. 2a involves
the $\Lambda N^{-1}$-$ \Lambda N^{-1}$ interaction whereas Figs. 2b can involve
both the $\Lambda N^{-1}$-$\Lambda N^{-1}$ and $NN^{-1}$-$NN^{-1}$
interactions.
In what follows we will not specifically consider the latter which involves no
strangeness flavor: We will assume it to be given by what is determined in
the non-strange (nuclear)
sector, {\it i.e.}, symmetry energy etc\footnote{In the sense of Fermi liquid,
this part of interactions should be given in terms of the standard
Landau-Migdal
interactions which we assume, as in condensed matter physics \cite{fixedpoint},
to be a fixed
point theory. As such, one can take this part of interactions to be
accurately given
by nuclear matter properties.}. Now for the s-wave in-medium kaon self-energy,
the relevant
four-Fermi interactions that involve a $\Lambda (1405)$ can be reduced to a
simple form involving two unknown constants
    \be
    {\cal L}_{4-fermion} &=& C_\Lambda^S \bar \Lambda \Lambda \Tr \bar B B
+  C_\Lambda^T \bar \Lambda \sigma^k \Lambda \Tr \bar B \sigma^k B
    \ee
where $B$ is the baryon (here nucleon) field,
$C_\Lambda^{S,T}$ are the dimension-2 ($M^{-2}$) parameters to be fixed
empirically and $\sigma^k$ acts on baryon spinor.

Additional (in-medium) two-loop graphs that involve $\Lambda N^{-1}$
excitations are given in Fig. 2c. They do not however involve
contact four-Fermi interactions, so are calculable unambiguously.

We shall denote the sum of these contributions from Figs. 2
to the self-energy by $\Pi _\Lambda$ \footnote{There is one class of
potentially
important in-medium two-loop graphs that we have not taken into account in this
paper and that is the set of graphs which could screen the leading
$\O (Q)$ term associated with the vector-meson ($\omega$ and $\rho$)
exchange \cite{LJMR}. To be specific, one
can visualize it as Fig. 1a with the box replaced by an
$\omega$-exchange vertex and with the nucleon loop attached to another nucleon
loop through a four-Fermi interaction. In the $\omega$-exchange channel,
the screening will go like
$\sim \frac{1}{1+F_0}$ where the $F_0$ is the Landau-Migdal parameter
\cite{pethick}. In dense matter, one expects $F_0 >0$.
Hence one might fear that there could be substantial
loss of attraction.  We believe this will not be serious here for two reasons.
Firstly the gauge coupling of the vector meson is predicted to scale {\em down}
as density increases as in the hidden-gauge symmetry Lagrangian theory
\cite{harada} so that $F_0$ will increase less rapidly than in the absence of
the scaling. Secondly the vector-exchange attraction is scaled {\em up} by the
BR scaling \cite{br91} by the factor $(m_\rho/m_\rho^\star)^2 >1$
where $m_\rho$ is the vector-meson mass with the star indicating the in-medium
quantity. These two opposing effects are expected to more or less cancel out.
We leave both effects out in this paper. They will be treated in detail
in a longer paper in preparation.}.

If one is only interested in critical density for kaon condensation and
in properties of kaonic atom, it suffices to compute the kaon self-energy. For
the equation of state to which we will return in a separate publication,
nonlinearities in the condensate fields will of course
have to be taken into account. This can be done in a straightforward way
as described in \cite{TPL}.
Putting all graphs up to two loops in medium, we have the complete
in-medium two-loop kaon self-energy
    \be
    \Pi_K(\omega) &=& -\left( \rho_p  {\cal T}^{K^-p}_{free}(\omega)
        +\rho_n  {\cal T}^{K^-n}_{free}(\omega) \right)
        - \left(\rho_p  \delta {\cal T}^{K^-p}_{\rho_N}(\omega)
        +\rho_n \delta {\cal T}^{K^-n}_{\rho_N}(\omega) \right)
        +\Pi_\Lambda(\omega)\label{self2}
    \ee
where ${\cal T}^{K^-N}_{free}$ is the scattering amplitude obtained in
LJMR \cite{LJMR}, $\rho_N$ ($N=p,n$) is the nucleon density
and $\delta{\cal T}^{K^-N}_{\rho_N}$ are
the medium modifications, by Figs.1b-1f, to ${\cal T}^{K^-N}_{free}$
\footnote{The explicit formulae will be given in a longer paper
in preparation.}
and
    \be
    \Pi_\Lambda(\omega) &=& - \frac{\bar g_\Lambda^2}{f^2}
        \bar g_\Lambda^2
        \left(\frac{\omega}{\omega+m_B-m_\Lambda} \right)^2
\left\{ C_\Lambda^S \rho_p \left( \rho_n +\frac 12 \rho_p \right)
    -\frac 32 C_\Lambda^T \rho_p^2 \right\}
\nonumber\\
&& -\frac{\bar g_\Lambda^4}{f^4} \rho_p
    \left(\frac{\omega}{\omega+m_B-m_\Lambda} \right)^2
    \omega^2 \left( \Sigma_K^p +\Sigma_K^n \right)
    \label{pilambda}\ee
where $\bar{g}_\Lambda$ is the renormalized $KN\Lambda (1405)$ coupling
constant determined in \cite{LJMR} and $\Sigma_K^N$ is given by,
\be
    \Sigma_i^{N}(\omega) &=&
      \frac{1}{2\pi^2} \int_0^{k_{F_N}} d|\vec k|
      \frac{|\vec k|^2}{\omega^2-m_i^2-|\vec k|^2}.
\ee
In eq.(\ref{pilambda}),
the first term comes from the diagrams of Figs. 2a and 2b
and the second term  from the diagrams of Figs. 2c.
While the second term gives repulsion corresponding to a Pauli quenching,
the first term can give either attraction or repulsion depending on
the sign of $(C_\Lambda^S [\rho_n+\frac 12 \rho_p]-
\frac 32 C_\Lambda^T\rho_p)$ with the constants $C_\Lambda^{S,T}$
being the only parameters that are not determined by
on-shell scattering data. We can fix them from kaonic
atom data \cite{kaonicatom} which require that there be an {\it effective
attraction}.
In condensed matter physics\cite{fixedpoint}, such an attractive four-Fermi
interaction in a particular kinematic situation turns by renormalization
into a ``marginally
relevant" interaction that causes instability of the system, {\eg},
pair condensation in superconductivity.
It is tempting to conjecture that something similar happens here, leading
to a strangeness condensation.
However what appears to be different here from a generic case in
condensed matter physics is that {\it the four-Fermi interaction contribution
(\ref{pilambda}) has a quadratic $\omega$ dependence which makes it
increasingly less effective toward the regime of small
$\omega$ where the condensation takes place.}

We have now all the ingredients needed to calculate the critical density.
For this, we will follow the procedure given in
\cite{TPL}. As argued in \cite{BKR}, we need not consider pions
when electrons with high chemical potential can trigger condensation through
the process $e^-\rightarrow K^- \nu_e$. Thus we can focus on the spatially
uniform condensate
    \be
    \langle K^-\rangle =v_K e^{-i\mu t}
    \ee
where $\mu$ is the chemical potential which is equal, by Baym's theorem
\cite{baym}, to the electron chemical potential. We shall parametrize
the proton and neutron densities by the proton fraction $x$
and the nucleon density $u=\rho/\rho_0$ as
    \be
    \rho_p = x\rho\; ,\;\; \rho_n =(1-x) \rho\; ,\;\; \rho = u\rho_0.
    \ee
Then the energy density $\tilde\epsilon$ -- which is related to the
effective potential in the standard way -- is given by,
    \be
    \tilde \epsilon (u,x,\mu, v_K) &=& \frac 35 E_F^{(0)} u^{\frac 53} \rho_0
        +V(u) +u\rho_0 (1-2x)^2 S(u) \nonumber\\
    &&-[\mu^2 -M_K^2 -\Pi_K (\mu,u,x)]
        v_K^2+{\cal O}(v_K^3)\nonumber\\
    && +\mu u\rho_0 x +\tilde\epsilon_e +\theta(|\mu|-m_\mu)\tilde \epsilon_\mu
    \ee
where $E_F^{(0)}=\left( p_F^{(0)}\right)^2/2m_N$ and $p_F^{(0)}=(3\pi^2\rho_0
/2)^{\frac 13}$ are, respectively,
 Fermi energy and momentum at nuclear density. The
$V(u)$ is a potential for symmetric nuclear matter
as described in \cite{PAL} which we believe is subsumed in contact four-Fermi
interactions (and one-pion-exchange -- nonlocal -- interaction)
in the non-strange sector as mentioned above. It will affect
the equation of
state in the condensed phase but not the critical density, so we will
drop it from now on. The nuclear symmetry energy $S(u)$ -- also
subsumed in four-Fermi interactions in the non-strange sector -- does
play a role as we know from ref.\cite{PAL}: Protons enter to neutralize the
charge of condensing $K^-$'s making the resulting compact star
``nuclear" rather than neutron star as one learns in standard astrophysics
textbooks. We take the form advocated in \cite{PAL}
    \be
    S(u) &=& \left(2^{\frac 23}-1\right) \frac 35 E_F^{(0)}
        \left(u^{\frac 23} -F(u) \right) +S_o F(u)
    \ee
where $F(u)$ is the potential contributions to the symmetry energy.
Three different forms of $F(u)$ were used in \cite{PAL}:
    \be
    F(u)=u\;,\;\; F(u) =\frac{2u^2}{1+u}\;,\;\; F(u)=\sqrt u.
\label{SE}
    \ee
It will turn out that
the choice of $F(u)$ does not significantly affect the critical density.

The contributions of the filled Fermi seas of electrons and muons are
\cite{TPL}
    \be
    \tilde \epsilon_e &=& -\frac{\mu^4}{12\pi^2} \nonumber\\
    \tilde \epsilon_\mu &=& \epsilon_\mu -\mu \rho_\mu \nonumber\\
    &=& \frac{m_\mu^4}{8\pi^2}\left((2t^2+1) t\sqrt{t^2+1}
-\ln(t^2+\sqrt{t^2+1}
    \right) -\mu \frac{p_{F_\mu}^3}{3\pi^2}
    \ee
where $p_{F_\mu} =\sqrt{\mu^2-m_\mu^2}$ is the Fermi momentum and $t=p_{F_\mu}
/m_\mu$.
The ground-state energy prior to kaon condensation
is obtained by extremizing $\tilde\epsilon$
with respect to  $x$, $\mu$ and $v_K$:
    \be
    \left. \frac{\partial\tilde\epsilon}{\partial x}\right |_{v_K=0}=0 \;,\;\;
    \left. \frac{\partial\tilde\epsilon}{\partial \mu}\right |_{v_K=0}=0
\;,\;\;
    \left. \frac{\partial\tilde\epsilon}{\partial v_K^2}\right |_{v_K=0}=0
    \ee
from which we obtain three equations corresponding, respectively,
to beta equilibrium, charge neutrality and dispersion relation:
    \be
    \mu &=& 4 (1-2 x) S(u) \nonumber\\
    0 &=& -x u\rho_0 +\frac{\mu^3}{3\pi^2}
         +\theta(\mu-m_\mu)\frac{p_\mu^3}{3\pi^2} \nonumber\\
    0 &=& D^{-1} (\mu, u,x) =\mu^2-M_K^2 -\Pi_K(\mu,u,x)\equiv \mu^2-
{M^\star_K}^2 (\mu,u,x)
    \ee
where $p_\mu =\sqrt{\mu^2-m_\mu^2}$. We have solved these equations using
for the kaon self-energy (a) the linear density approximation, eq.(\ref{self1})
and (b) the full two-loop result, eq.(\ref{self2}). Table 1 shows the case (a)
for different symmetry energies eq.(\ref{SE}). We see that the precise form of
the symmetry energy does not matter quantitatively, so we will simply take
$F(u)=u$ from now on. The corresponding ``effective kaon mass" ${M^\star_K}$
is plotted vs. $u$ in Fig. 3 in solid line. Note that even in this linear
density approximation kaon condensation {\it does} take place, {\it albeit} at
a bit higher density than obtained before.
    \begin{table}
    $$
    \begin{array}{|c|c|c|c|}
    \hline
     F(u) &  u & \frac{2 u^2}{1+u} & \sqrt u \\
    \hline
    u_c   & 3.90 & 3.77 & 4.11 \\
    \hline
    \end{array}
    $$
\vskip 0.2cm
    \centerline{Table 1 : Critical density in linear density approximation
                corresponding to Fig. 1a}
    \end{table}

We now turn to the (in-medium) two-loop calculation. For this we need to fix
the parameters $C_\Lambda^{S,T}$. This could be done with kaonic atom data.
The presently available data \cite{kaonicatom} imply that the optical
potential for the $K^-$ in medium has an attraction of the order of
    \be
    \Delta V \approx -(180\pm 20)\ \ {\rm  MeV}\ \ at \;\; u=0.97
    \ee
which implies approximately for $x=1/2$
\be
 ( C_\Lambda^S -C_\Lambda^T) f_\pi^2\approx 10.\label{cvalue}
\ee
This leaves one parameter free for $x\neq 1/2$. It will, however, turn out
that this
freedom does not diminish significantly the predictiveness of the theory.
We shall not attempt to fine-tune these constants in this work but
take (\ref{cvalue}) to be some mean value in what follows.
The result turns out be pretty much insensitive to the precise values
of the constants.
When one can pin them down (through, {\it e.g.}, isotope effects)
from kaonic atom data, our prediction could be made considerably more precise.

In Table 2, we list the predicted density dependence of the real part of
the kaonic atom potential for $x=0.5$ obtained for
$(C_\Lambda^S -C_\Lambda^T) f_\pi^2 \approx 10$.
To exhibit the role of $\Lambda(1405)$ in the kaon self-energy, we list
each contribution of $\Pi$. Here $\Pi_{free}=
-\rho_N {\cal T}^{K^-N}_{free}$, $\delta\Pi = -\rho_N
\delta {\cal T}^{K^-N}$, $\Pi_\Lambda^1$  corresponds to the first term
of eq.(\ref{pilambda}) which depends on $C_\Lambda^{S,T}$  and $\Pi_\Lambda^2$
to the second term independent of $C_\Lambda^{S,T}$.
We observe that the $C_\Lambda^{S,T}$-dependent term plays a crucial role
for attraction in kaonic atom.
\begin{table}
$$
\begin{array}{|c|c|c|c|c|c|c|c|}
\hline
 u & M_K^* & \Delta V & x & -\rho {\cal T}^{free} & -\rho \delta {\cal
T}^{free}
&
\Pi^1_\Lambda & \Pi^2_\Lambda \\
\hline
 0.2& 424.6& -70.37& 0.5& -0.0673& 0.0034& -0.1998& 0.007607\\
 0.4& 390.0& -105.0& 0.5& -0.0920& 0.0084& -0.3024& 0.006641\\
 0.6& 364.3& -130.7& 0.5& -0.1173& 0.0143& -0.3610& 0.005635\\
 0.8& 342.6& -152.4& 0.5& -0.1462& 0.0207& -0.3996& 0.004794\\
 1.0& 323.5& -171.5& 0.5& -0.1789& 0.0274& -0.4250& 0.004088\\
 1.2& 306.2& -188.8& 0.5& -0.2150& 0.0343& -0.4404& 0.003483\\
 1.4& 289.9& -205.1& 0.5& -0.2540& 0.0413& -0.4460& 0.002945\\
\hline
\end{array}
$$
\vskip 2mm
\begin{center}
\parbox{5.5in}{
 Table 2 : Self-energies for kaonic atom in nuclear matter ($x=0.5$)
in unit of $M_K^2$ for $(C_\Lambda^S-C_\Lambda^T) f_\pi^2 =10$ and $F(u)=u$.
$\Delta V\equiv M_K^\star -M_K$ is the attraction (in unit of MeV) at
given density.}
\end{center}
\end{table}
    \begin{table}
    $$
    \begin{array}{|c|c|c|c|}
    \hline
C_\Lambda^S f_\pi^2 & 10 & 5 & 0 \\
    \hline
 u_c & 3.13 & 3.33 & 3.69 \\
    \hline
    \end{array}
    $$
\vskip 0.2cm
\begin{center}
 \parbox{4in}{Table 3 : Critical density $u_c$ in in-medium two-loop chiral
perturbation theory for $(C_\Lambda^S-C_\Lambda^T) f_\pi^2=10$ and $F(u)=u$.}
\end{center}
    \end{table}
For the value that seems to be required by the kaonic atom
data, (\ref{cvalue}),
the critical density comes out to be about $u_c\approx 3$, rather close
to the original Kaplan-Nelson value.

In Table 3 and Fig. 3 are given the predictions for a wide range of
values for $C_\Lambda^S f_\pi^2$. What is remarkable here is
that while the $C_\Lambda^{S,T}$-dependent four-Fermi interactions are
{\it essential} for triggering kaon condensation,
the critical density is quite insensitive to their strengths.
In fact, reducing the constant $(C_\Lambda^S-C_\Lambda^T)f_\pi^2$
that represents the kaonic atom attraction
by an order of magnitude to 1 with
$C_\Lambda^S f_\pi^2=10,\ 0$ modifies the
critical density only to $u_c\approx 3.3,\ 4.5$, respectively.

In conclusion, we have shown that chiral perturbation theory at
order N$^2$LO predicts kaon condensation in ``nuclear
star" matter at a density $\rho_c <4 \rho_0$ with
a large fraction of protons -- $x=0.1\sim 0.2$ at the critical point and
rapidly increasing afterwards -- neutralizing the negative charge of
the condensed kaons. For this to occur, four-Fermi interactions
involving $\Lambda (1405)$ are found to play an important role in triggering
the
condensation but the critical density is surprisingly insensitive to the
strength of the four-Fermi interaction. This makes the condensation phenomenon
even more robust than thought before.
We have not taken into account the BR scaling \cite{br91} -- which we should,
to be fully consistent. An approximate account of the BR scaling is found to
lower the critical density below $3 \rho_0$, practically
independently of the magnitude
of $C_\Lambda^{S,T}$. As remarked above, there are compensating effects such as
the screening of the vector channel which has to be considered on the same
footing, so a fully consistent treatment would require more work.
What is fairly certain is that although the detailed mechanism appears
quite different here from that in the low-order treatments,
once the condensation sets in, the rest of the star properties
are expected to resemble closely the structure obtained in
\cite{TPL,BLRT}. A detailed analysis of the ``nuclear
star" that we obtained in this paper will be made elsewhere.

\vskip 0.5cm
\centerline{\large \bf Acknowledgments}
CHL is grateful for valuable discussions with H. Jung and D.-P. Min.
He would also like to thank V. Thorsson for informative
discussions at the 1993 Chiral
Symmetry Workshop at ECT$^*$ in Trento. We have benefited from conversations
with N. Kaiser, K. Kubodera, A. Manohar and H. Yabu. Part of this work was done
while two of us (CHL and MR) were attending the 94 Winter Workshop in
Daejeon (Korea) sponsored by the Center for Theoretical Physics of Seoul
National University.
The work of CHL was supported in part by the Korea Science and
Engineering Foundation through the CTP of SNU and in part by the
Korea Ministry of Education and
the work of GEB by the US Department of Energy under Grant
No. DE-FG02-88ER40388.

\vskip 5mm
\hrule
\vskip 5mm

\centerline{\bf Figure Captions}
\begin{itemize}
\item {\bf Figure 1:}
(a): The linear density approximation to the kaon self-energy in medium,
$\Pi_K$.
The square blob represents the off-shell $K^-N$ amplitude
calculated to $\O (Q^3)$;
(b)-(f): medium corrections to ${\cal T}^{KN}$ of fig.(a) with the free nucleon
propagator indicated by a double slash replaced by an in-medium one,
eq.(\ref{prop}). The loop labeled $\rho_N$ represents the in-medium nucleon
loop
proportional to density, $N^{-1}$ the nucleon hole ($n^{-1}$ and/or $p^{-1}$),
the external dotted line stands for
the $K^-$ and the internal dotted line for the pseudoscalar octets $\pi$,
$\eta$, $K$.
\item {\bf Figure 2:}
Two-loop diagrams involving $\Lambda(1405)$ contributing
to the kaon self-energy. Diagrams (a)
and (b) involve four-Fermi interactions describing the $\Lambda
N^{-1}$-$\Lambda
N^{-1}$ vertex. Diagram (c) does not involve four-Fermi interactions
and hence is unambiguously determined by on-shell parameters. Here the
internal dotted line represents the kaon.
\item {\bf Figure 3:}
Plot of the effective kaon mass $M^\star_K$ obtained from
the dispersion formula $D^{-1}(\mu,u)$ $=0$ vs. the chemical
potential $\mu$ prior to kaon condensation, with
$F(u)=u$. The solid line corresponds to the linear density approximation and
the dashed lines to the in-medium two-loop results for
$(C_\Lambda^S-C_\Lambda^T) f_\pi^2 =10$ and
$C_\Lambda^S  f_\pi^2 = 10, 5, 0$ respectively from the left.
The point at which  the chemical potential $\mu$ (denoted $\mu_{cp}$)
intersects
$M_K^\star$ corresponds to the critical point.
\end{itemize}

\end{document}